# Large-scale analysis of disease pathways in the human interactome

Monica Agrawal[1,‡], Marinka Zitnik[1,‡], and Jure Leskovec[1,2]
[1]*Department of Computer Science, Stanford University, Stanford, CA, USA*
[2]*Chan Zuckerberg Biohub, San Francisco, CA, USA*
[‡]*Equal contribution; Email: {agrawalm, marinka, jure}@cs.stanford.edu*

Discovering disease pathways, which can be defined as sets of proteins associated with a given disease, is an important problem that has the potential to provide clinically actionable insights for disease diagnosis, prognosis, and treatment. Computational methods aid the discovery by relying on protein-protein interaction (PPI) networks. They start with a few known disease-associated proteins and aim to find the rest of the pathway by exploring the PPI network around the known disease proteins. However, the success of such methods has been limited, and failure cases have not been well understood. Here we study the PPI network structure of 519 disease pathways. We find that 90% of pathways do not correspond to single well-connected components in the PPI network. Instead, proteins associated with a single disease tend to form many separate connected components/regions in the network. We then evaluate state-of-the-art disease pathway discovery methods and show that their performance is especially poor on diseases with disconnected pathways. Thus, we conclude that network connectivity structure alone may not be sufficient for disease pathway discovery. However, we show that higher-order network structures, such as small subgraphs of the pathway, provide a promising direction for the development of new methods.

## 1. Introduction

Computational discovery of disease pathways aims to identify proteins and other molecules associated with the disease.[1–3] Discovered pathways are systems of interacting proteins and molecules that, when mutated or otherwise altered in the cell, manifest themselves as distinct disease phenotypes (Figure 1A).[4] Disease pathways have the power to illuminate molecular mechanisms but their discovery is a challenging computational task. It involves identifying all disease-associated proteins,[2,5,6] grouping the proteins into a pathway,[7–10] and analyzing how the pathway is connected to the disease at molecular and clinical levels.[11,12] Many of the main challenges facing the task arise from the interconnectivity of a pathway's constituent proteins.[2,13–15] This interconnectivity implies that the impact of altering one protein is not restricted only to the altered protein, but can spread along the links of the protein-protein interaction (PPI) network[14] and affect the activity of proteins in the vicinity.[4,15]

As understanding each disease protein in isolation cannot fully explain most human diseases, numerous computational methods were developed to predict which proteins are associated with a given disease, and to bring them together into pathways using the PPI network (Figure 1B).[2,5–10,16,17] These methods have accelerated the understanding of diseases, but have not yet fully succeeded in providing actionable knowledge about them.[1] For example, recent studies[5,7,18] found that only a relatively small fraction of disease-associated proteins physically interact with each other, suggesting that methods, which predict disease proteins by searching for dense clusters/communities of interacting proteins in the network, may be limited in discovering disease pathways. Analytic methods may thus be hindered by such issues, and unless specifically tuned, can lead to an expensive and time-consuming hunt for new disease proteins. Furthermore, although numerous methods exist, protein-protein interaction and connectivity patterns of disease-associated proteins remain largely unexplored.[4,12] Because of the huge potential of these methods for the development of better strategies for disease prevention, diagnosis, and treatment, it is thus critical to identify broad conceptual and methodological limitations of current approaches.

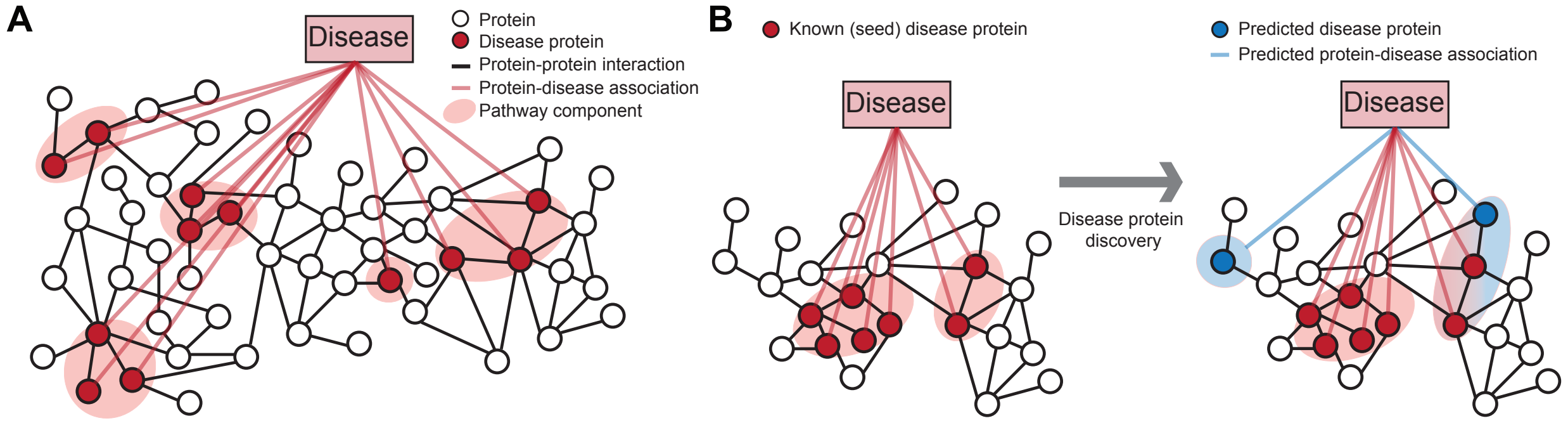

Fig. 1. **Network-based discovery of disease proteins. A** Proteins associated with a disease are projected onto the protein-protein interaction (PPI) network. In this work, *disease pathway* denotes a (undirected) subgraph of the PPI network defined by the set of disease-associated proteins. The highlighted disease pathway consists of five pathway components. **B** Methods for disease protein discovery predict candidate disease proteins using the PPI network and known proteins associated with a specific disease. Predicted disease proteins can be grouped into a disease pathway to study molecular disease mechanisms.

**Present work.** Here we study the PPI network structure of 519 diseases. For each disease, we consider the associated disease proteins and project them onto the PPI network to obtain the disease pathway (Figure 1A).[1,4] We then investigate the network structure of these disease pathways.[a]

We show that disease pathways are fragmented in the PPI network with on average more than 16 disconnected pathway components per disease. Furthermore, we find that each component contains only a small fraction of all proteins associated with the disease. Through spatial analysis of the PPI network we find that proximity of disease-associated proteins within the PPI network is statistically insignificant for 92% (476 of 519) diseases, and that 90% of diseases are associated with proteins that tend not to significantly interact with each other, indicating that disease proteins are weakly embedded—rather than densely interconnected—in the PPI network.

We then consider state-of-the-art network-based methods for disease protein discovery (Figure 1B). These methods use the PPI network and a small set of known disease proteins to predict new proteins that are likely associated with a given disease. However, as we show here, current methods disregard loosely connected proteins when making predictions, causing many disease pathway components in the PPI network to remain unnoticed. In particular, we find that performance of present methods is better for diseases whose pathways have high edge density, are primarily contained within a single pathway component, and are proximal in the PPI network. However, our analysis shows that a vast majority of disease pathways does not display these characteristics.

The search for a solution to the better characterization of disease pathways has led us to study higher-order protein-protein interaction patterns[4,19–21] of disease proteins. Following on from earlier work[22] showing that higher-order PPI network structure around cancer proteins is different from the structure around non-cancer proteins, we find that many proteins associated with the same disease are involved in similar higher-order network patterns, even if disease proteins are not adjacent in the PPI network. In particular, we find that proteins associated with 60% (310 of 519) of diseases do exhibit over-representation for certain higher-order network patterns, suggesting that disease proteins can take on similar structural roles, albeit located in different parts of the PPI network. We demonstrate that taking these higher-order network structures into account can shrink the gap between current and goal performance of disease protein discovery methods.

[a]All data and supplementary tables with results are at: `http://snap.stanford.edu/pathways`.

In addition to new insights into the PPI network connectivity of disease proteins, our analysis on network fragmentation of disease proteins and their distinctive higher-order PPI network structure leads to important implications for future disease protein discovery that can be summarized as:

- We move away from modeling disease pathways as highly interlinked regions in the PPI network to modeling them as loosely interlinked and multi-regional objects with two or more regions distributed throughout the PPI network.
- Higher-order connectivity structure provides a promising direction for disease pathway discovery.

## 2. Background and related work

Next, we give background on disease pathways and on methods for disease protein prediction.

**Disease pathways.** Broadly, a disease pathway in the PPI network is a system of interacting proteins whose atypical activity collectively produces some disease phenotype.[3,4,12,16] Given the PPI network $G = (V, E)$, whose nodes $V$ represent proteins and edges $E$ denote protein-protein interactions, the *disease pathway* for disease $d$ is an undirected subgraph $H_d = (V_d, E_d)$ of the PPI network specified by the set of proteins $V_d$ that are associated with $d$, and by the set of protein-protein interactions $E_d = \{(u, v)|(u, v) \in E \text{ and } u, v \in V_d\}$ (*e.g.*, Adrenal cortex carcinoma pathway in Figure 4). To measure the specifics of protein interactions within and outside the pathway we define pathway boundary as the set $B_d = \{(u, v)|(u, v) \in E, u \in V_d, v \in V \setminus V_d\}$ consisting of all edges that have one endpoint inside $H_d$ and the other endpoint outside $H_d$.

**Network-based methods for disease protein discovery.** Given a specific disease, the task is to take the PPI network and the disease proteins and to predict new proteins that are likely associated with the disease. Approaches for this task are known as protein-disease association prediction or disease module detection methods (Figure 1B), and can be grouped intro three categories. (1) Neighborhood scoring and clustering methods[4,5,7,9,10,12] assume that proteins that belong to the same network cluster/community are likely involved in the same disease. In direct neighborhood scoring, each protein is assigned a score that is proportional to the percentage of its neighbors associated with the disease. To identify clusters that extend beyond direct neighbors, the methods start with a small set of disease proteins (seed proteins) and grow a cluster by expanding the seeds with the highest scoring proteins. However, few existing methods (*e.g.*, connectivity significance-based method DIAMOnD) can work with seed proteins that are not adjacent in the PPI network.[7] (2) Diffusion-based methods[5,8,23] use seed proteins to specify a random walker that starts at a particular seed protein and at every time step moves to a randomly selected neighbor protein. Upon convergence, the frequency with which the nodes in the network are visited is used to rank the corresponding proteins. (3) Representation learning methods,[6,16,17,21,24] such as matrix completion, graphlet degree signatures, and neural embeddings, construct representations for proteins (*i.e.*, latent factors, embeddings) that capture known protein-disease associations and/or proteins' network neighborhoods, and then use these representations as input to a downstream predictor. We consider a neural embedding approach[24] that first learns a vector representation for each protein using a single-layer neural network and random walks and then fits a logistic regression classifier that predicts disease proteins based on these feature vectors. We also consider a matrix completion method[6] that factorizes a protein-disease association matrix into a set of protein and a set of disease latent factors while also incorporating the PPI network. Predictions for a new disease are obtained as a function of the feature and latent factors.

Although many methods exist for predicting disease proteins, surprisingly little is known about the PPI network structure of disease pathways and how it relates to the power of these methods.

## 3. Data

We continue by describing the datasets used in this study.

**Human protein-protein interaction network.** We use the human PPI network compiled by Menche *et al.*[18] and Chatr-Aryamontri *et al.*.[25] Culled from 15 databases, the network contains physical interactions experimentally documented in humans, such as metabolic enzyme-coupled interactions and signaling interactions. The network is unweighted and undirected with $n = 21{,}557$ proteins and $m = 342{,}353$ experimentally validated physical interactions. Proteins are mapped to genes and the largest connected component of the PPI network is used in the analysis. We also investigate two other PPI network datasets to make sure that our findings are not specific to the version of the PPI network we are using. Unless specified, results in the paper are stated with respect to the first dataset. The other two PPI networks are from the BioGRID database[25] and the STRING database.[26] Both of these networks are restricted to those edges that have been experimentally verified.

**Protein-disease associations.** A protein-disease association is a tuple $(u, d)$ indicating that alteration of protein $u$ is linked to disease $d$. Protein-disease associations are pulled from DisGeNET, a platform that centralized the knowledge on Mendelian and complex diseases.[2] We examine over 21,000 protein-disease associations, which are split among the 519 diseases that each has at least 10 disease proteins. The diseases range greatly in complexity and scope; the median number of associations per disease is 21, but the more complex diseases, *e.g.*, cancers, have hundreds of associations.

**Disease categories.** Diseases are subdivided into categories and subcategories using the Disease Ontology.[27] The diseases in the ontology are each mapped to one or more Unified Medical Language System (UMLS) codes, and of the 519 diseases pulled from DisGeNET, 290 have a UMLS code that maps to one of the codes in the ontology. For the purposes of this study, we examine the second-level of the ontology; this level consists of 10 categories, such as cancers (68 diseases), nervous system diseases (44), cardiovascular system diseases (33), and immune system diseases (21).

Altogether, we use human disease and PPI network information that is more comprehensive than in previous works,[7,18,22] which focused on smaller sets of diseases and proteins.

## 4. Connectivity of disease proteins in the PPI network

We start by examining the network connectivity of disease proteins. We then analyze disease protein discovery methods and contextualize their performance using disease pathway network structure.

### 4.1. *Proximity of disease proteins in the PPI network*

We begin by briefly describing network measures that we use to characterize connectivity of disease proteins, both within disease pathways and with respect to the rest of proteins in the PPI network.

**PPI network distance and concentration measures.** We consider the following measures to characterize PPI connectivity of disease proteins for each disease $d$ and its associated pathway $H_d$:

- *Size of largest pathway component:* Fraction of disease proteins that lie in $H_d$'s largest pathway component (*i.e.*, the relative size of the largest connected component (LCC) of $H_d$).

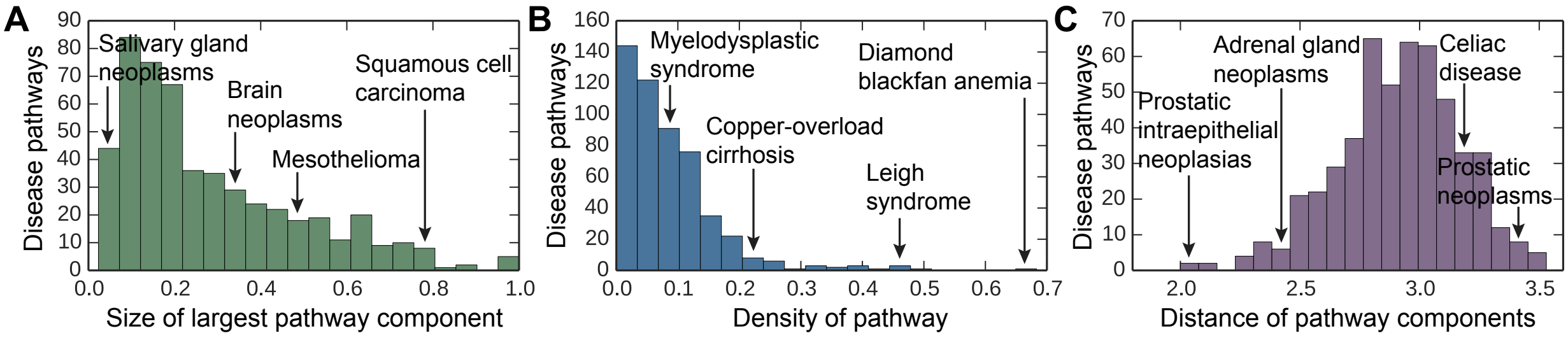

Fig. 2. **Protein interaction connectivity of disease pathways.** The distribution of (**A**) the network densities of each disease pathway, (**B**) the relative size of the largest pathway component calculated as a fraction of disease proteins that lie in the largest pathway component, and (**C**) the average shortest path length between disparate pathway components in the PPI network.

- *Density of the pathway:* It is calculated as: $2|E_d|/(|V_d|(|V_d|-1))$ and takes values in $[0,1]$. A higher density indicates a higher fraction of edges (out of all possible edges) appear between nodes in $H_d$.
- *Distance of pathway components:* For each pair of pathway components (Figure 1A), we calculate the average shortest path length between each set of proteins, and then, the average of this is taken over all pairs of the components.
- *Conductance:*[28] It is calculated as: $|B_d|/(|B_d|+2|E_d|)$ and takes values in $[0,1]$. A lower conductance indicates the pathway is a more well-knit community separated from the rest of the network.
- *Spatial network association:*[29,30] It measures concentration/localization of disease proteins in the PPI network by quantifying how strongly disease proteins co-cluster within the PPI network and whether this co-clustering is stronger than expected by random chance. It is calculated as: $K_d(s) = 2/(\bar{p}n)^2 \sum_i p_i \sum_j (p_j - \bar{p}) I(\ell_G(i,j) < s)$, where $p_i$ is a binary indicator indicating if node $i$ represents a $d$-associated protein, $\bar{p} = 1/n \sum_i p_i$, and $I(\ell_G(i,j) < s)$ equals 1 if the shortest path length between $i$ and $j$ is less than $s$ and 0 otherwise. If all disease proteins lie in one PPI network region, most of them are found for small values of $s$, while for uniformly spread proteins in the PPI network $K_d(s)$ achieves larger values only for large values of $s$. The significance of $H_d$'s concentration is determined by computing the area under the $K_d(s)$ curve[29] for $d$-associated proteins and comparing it to curves obtained by applying the same statistic to sets of random proteins.
- *Network modularity:*[31] Fraction of edges that fall within/outside the pathway minus the expected fraction if edges were randomly distributed: $Q_d = 1/(2m)\sum_{ij}(I((i,j)\in E) - \frac{k_i k_j}{2m})\delta(p_i,p_j)$, where $k_i$ is the degree of $i$, and $\delta(p_i,p_j)$ is 1 if $p_i$ and $p_j$ are equal and 0 otherwise.

**PPI network structure of disease pathways.** First, we find that disease pathways are fragmented in the PPI network, with a median of 16 connected components per disease and a median of only 21% of the proteins lying in the largest pathway component (Figure 2A). Only approximately 10% of pathways have over 60% of their proteins in the largest pathway component. We also find that disease pathways are not particularly well connected internally with only a median density of 0.07 (the overall PPI network density is 0.0015), and 90% of diseases have a density below 0.17 (Figure 2B). Furthermore, they are rather well connected externally, having a median conductance of 0.96, meaning that the disease pathway has relatively as many edges pointing outside the pathway to the rest of the PPI network as it has edges lying inside the pathway. The median distance between the pathway components is almost 2.9 (Figure 2C). These results counter expectations as they show that disease pathways do not have the PPI network structure one expects of a traditional network cluster/community, which is well connected internally and has few edges pointing outside the cluster.[14,31]

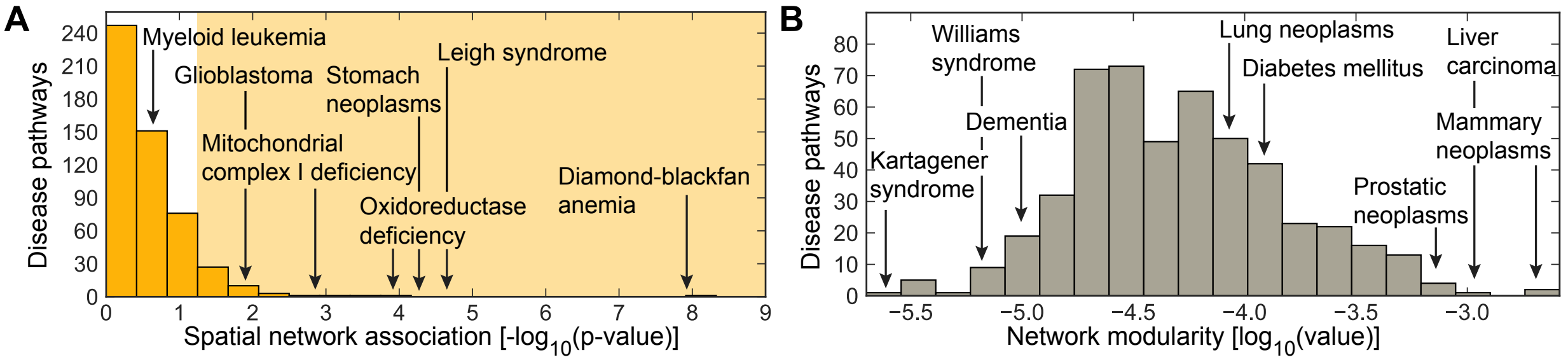

Fig. 3. **Spatial clustering and modular structure of disease pathways in the PPI network.** The distribution of (**A**) the spatial clustering calculated for each disease pathway as the strength of association[29] between the set of disease proteins and the PPI network (shaded area indicates significant spatial clustering at $\alpha = 0.05$ level), and (**B**) the modularity[31] of disease pathways in the PPI network.

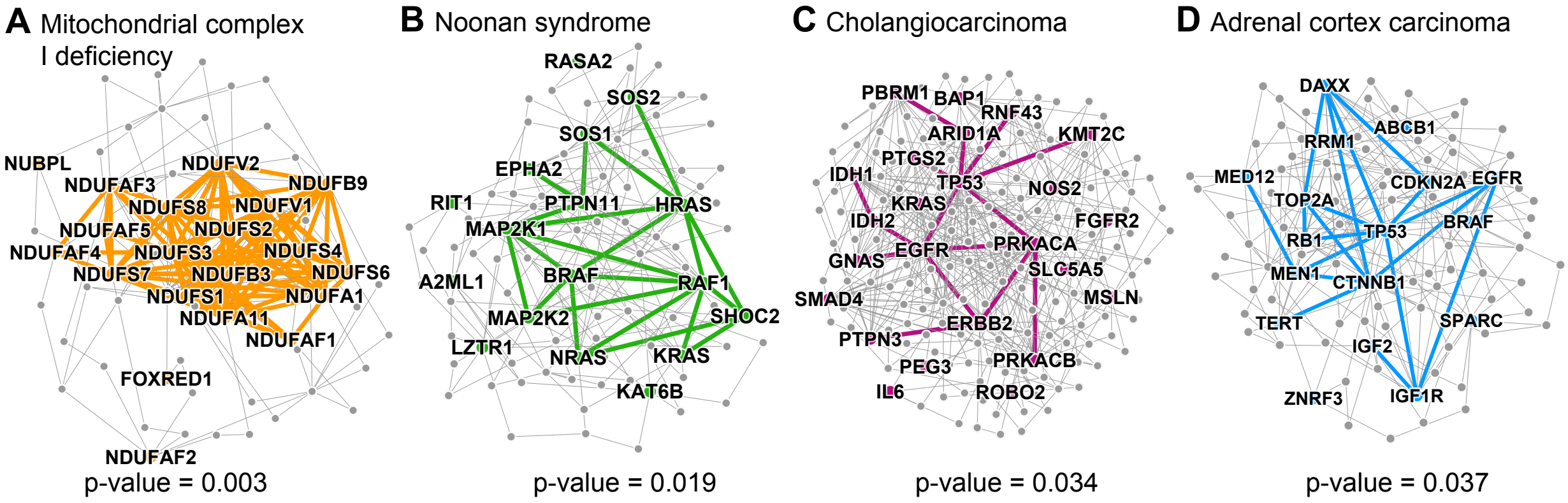

Fig. 4. **Disease pathways in the wider PPI network**. A small PPI subnetwork highlighting physical interactions between disease proteins associated with (**A**) Mitochondrial complex I deficiency, (**B**) Noonan syndrome, (**C**) Cholangiocarcinoma, and (**D**) Adrenal cortex carcinoma. Shown are selected disease pathways whose spatial clustering[29] within the PPI network is statistically significant (p-values shown; entire distribution of the p-values is shown in Figure 3A) and is also among the strongest (top-30 diseases) in the disease corpus.

To statistically test how well disease pathways are localized in the PPI network, we conduct spatial analysis of the PPI network. We find no significant pathway localization for 92% of diseases (Figure 3A), suggesting that these diseases have pathways that are multi-regional with two or more regions of disease proteins in different parts of the PPI network. The presence of multiple regions suggests that each disease might be comprised of several groups of proteins that are located in weakly connected or disconnected regions of the PPI network and thus may be functionally distinct.[1,4] We find that only 43 of 519 (8%) diseases are region-specific (examples in Figure 4), *i.e.*, they significantly associate with only one local neighborhood and can be found in a single region of the PPI network. We also observe that the number of edges within a disease pathway rarely exceeds the number expected on the basis of chance (Figure 3B). The median modularity of disease pathways is only $4.6 \times 10^{-4}$, reflecting there is no significant concentration of edges within disease pathways compared with random distribution of edges between all proteins regardless of pathways. These results suggest that integration of disconnected regions of disease proteins into a broader disease pathway will be crucial for a holistic understanding of disease mechanisms.

Finally, we note that these findings can be reproduced in three PPI network datasets (Section 3),

suggesting that our key results are robust against potential biases in the PPI network data.

### 4.2. *Connections between PPI network structure and disease protein discovery*

Next, we study disease protein discovery methods based on some of the most frequently used principles for identifying disease proteins,[14] and evaluate them through PPI pathway network structure.

**Methods and experimental setup.** We consider five methods: direct neighborhood scoring,[5] neural embeddings,[24] matrix completion,[6] network diffusion,[8,23] and connectivity significance (DIAMOnD).[7] See Section 2 for details on the methods. We use disease-centric ten-fold cross-validation. For each disease, the set of all proteins is randomly split into ten folds, with each fold containing an equal number of proteins associated with that disease. In each of the ten runs, the goal is to predict disease proteins in the test fold, assuming knowledge of disease proteins in the nine other folds. Each method assigns a score to each protein in the network representing the probability that the protein is associated with the disease. 20 diseases are set aside for hyperparameter selection, and the remaining 499 are used for testing. For evaluation, recall-at-$k$ is measured to quantify what fraction of all the disease proteins are ranked within the first $k$ predicted proteins (*e.g.*, $k = 25, 100$). We also calculate the mean reciprocal rank (MRR) for all of the algorithms in order to get an overall measure of method performance. Measures range between 0 and 1, and a higher score indicates better performance.

**Prediction performance in the context of disease pathway structure.** Figure 5 shows performance of disease protein discovery methods as a function of PPI connectivity of disease proteins. We observe that the higher the degree of agglomeration of disease proteins within the PPI network, the higher the performance of prediction methods. In particular, across all five methods, performance correlates positively with density and percent of proteins in the largest pathway component, and negatively with the distance between pathway components. These correlations are weaker for the Neural embeddings than for the other four methods, but the overall direction of the trends is the same across all of them. For example, the correlation between density and recall-at-100 is $\rho = 0.45$ for the Neural embeddings and between $\rho = 0.54$ and $\rho = 0.63$ for the other four methods (Figure 5B).

Across all diseases, random walk-based methods are the best performers, as evaluated by both mean reciprocal rank (MRR = 0.061 and MRR = 0.050 for Random walk and Neural embeddings, respectively) and recall (Recall-at-100 = 0.356 and Recall-at-100 = 0.300 for Random walk and Neural embeddings, respectively). However, we see that Random walk method is particularly dependent on the percent of disease proteins in the largest pathway component (Figure 5A). The difference in recall between the Random walk method and the Neural embeddings is positively correlated ($\rho =$ 0.26) with that percentage. Since random walks and other diffusion-based variants are very popular, it is problematic that they are reliant on properties that are not typical of disease pathways.

Neighborhood scoring performs the worst by both metrics (MRR = 0.029, Recall-at-100 = 0.242). The superior performance of random walk-based methods indicates that the assumption that the neighborhood method makes in calculating scores based only on protein's neighbors is too restrictive when defining disease locality.[5] Though DIAMOnD does not outperform Random walk, we observe that it has a comparable recall in its higher-ranked predictions (recall-at-25 = 0.186, compared to Random walk's 0.199), but its performance lags considerably for lower-ranked predictions (Recall-at-100 = 0.300, compared to Random walk's 0.356).

We see the most complementarity between Neural embeddings, Matrix completion and the other methods, which makes sense given that the other three methods are all based on direct/indirect network neighborhood scoring, while Neural embeddings and Matrix completion more flexibly capture network structure and neighborhoods of disease proteins.[6] For example, we can examine the disease pathway for Juvenile myelomonocytic leukemia in which Neural embeddings method performs far better than Random walk (Recall-at-100 = 0.550, compared to Random walk's 0.200). The pathway consists of nineteen nodes, but there exists only three edges within the pathway (network density = 0.008). Therefore, the Neural embeddings method is able to capture latent features about pertinent nearby nodes that Random walk struggles to find, given that Random walk is highly dependent on the edges near the seed proteins. On the other hand, the pathway for Squamous cell carcinoma is more accurately detected by Random walk than by Neural embeddings (Recall-at-100 = 0.540, compared to Neural embeddings' 0.120). This can be explained by higher interconnectivity of Squamous cell carcinoma pathway in the PPI network (network density = 0.034) suggesting that there are advantages to focusing on local edge connectivity.

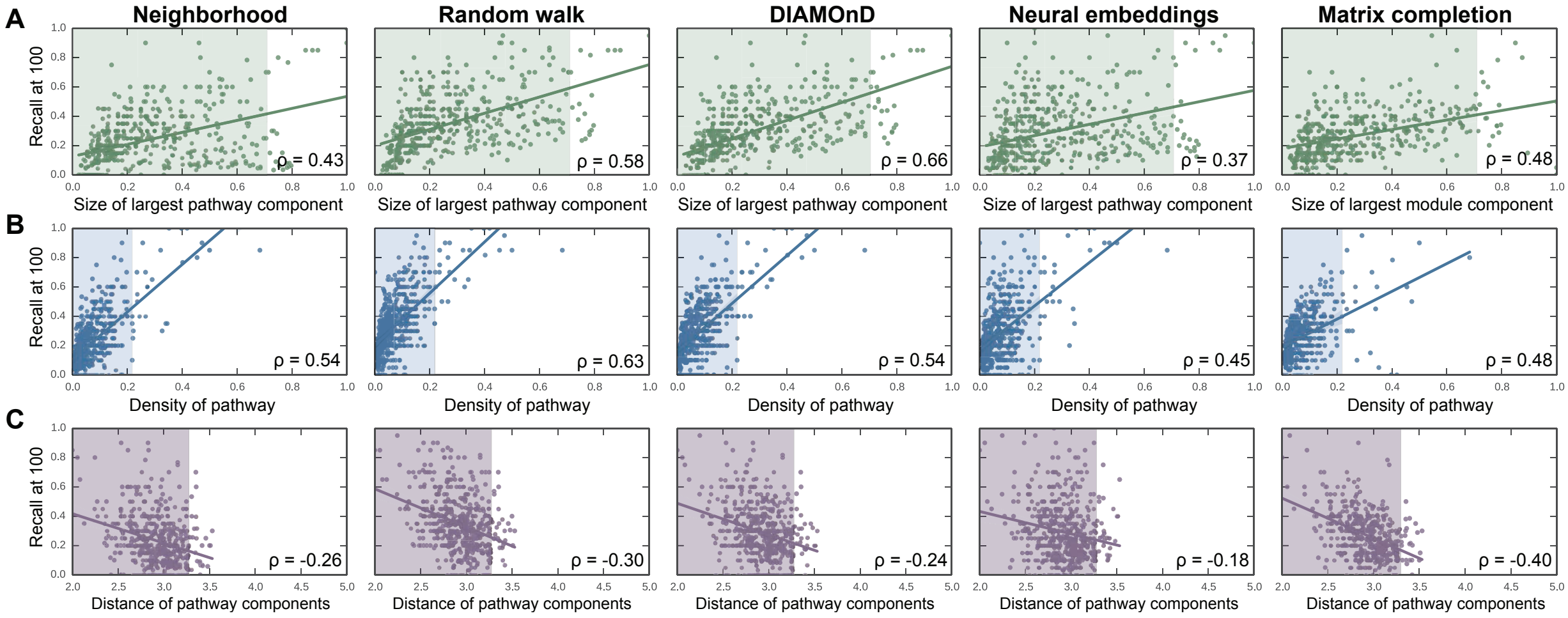

Fig. 5. **Prediction quality versus PPI connectivity of disease proteins.** Each point represents one disease; its location is determined by the quality of predicted disease proteins (y-coordinate), and by the connectivity of disease proteins in the PPI network (x-coordinate). Across all five methods, the trends uniformly indicate that (**A**) the bigger the largest pathway component, (**B**) the more densely interconnected the disease pathway, and (**C**) the lower the average shortest path length between disparate pathway components, the better the predictions. The shaded areas represent the space in which 95% (494 of 519) of all diseases reside.

**Performance variation across disease categories.** We observe strong differences in performance across disease categories indicating that diseases should not be considered as a monolithic category, given the very different mechanisms behind them. The same performance patterns hold across all five of the methods, suggesting that none of the assumptions each method makes about pathway structure correspond to any of the particular mechanisms that tend to be more specific to one disease category. Furthermore, because of the similar performance among methods over easy[32] (*e.g.*, median recall-at-100 = 0.720 for Mendelian diseases) and difficult[32] (*e.g.*, median recall-at-100 = 0.360 for cancer diseases) disease categories, the assumptions made by current methods do not seem to accurately reflect the uncertainty associated with a protein's true association with a disease.

## 5. Higher-order connectivity of disease proteins in the PPI network

We showed in Section 4 that proximity of disease proteins in the PPI network is likely insufficient for the disease protein discovery task as disease pathways have rather low PPI density and a rather high conductance. To look past just edge connectivity for the prediction of disease proteins, we in-vestigate what higher-order PPI network structures disease proteins are likely to be involved in, and then incorporate this structure information to augment prediction capability of current methods.

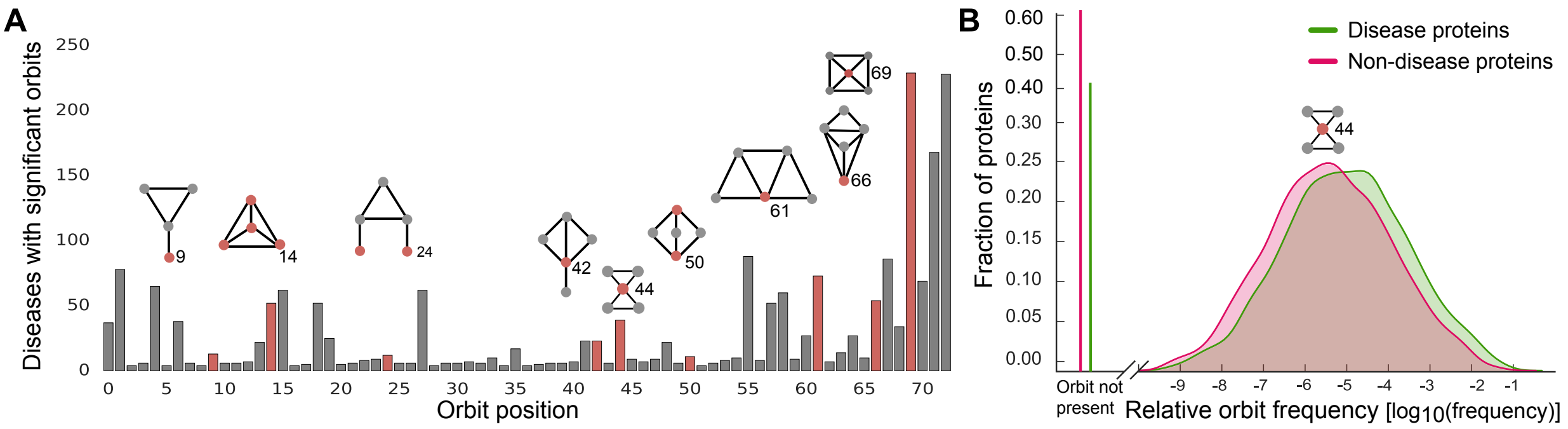

Fig. 6. **Over-representation of motifs in disease modules. A** The number of diseases (out of 519 possible) for which the associated proteins are significantly over-represented at each orbit position. A disease is deemed significant at a given orbit position if the median number of times a disease protein matching that position was significant at $\alpha = 0.01$, as compared to permutation testing over random sets of proteins of the same size. Pictured above are selected motifs (red node represents the orbit position, *i.e.*, the location where the node touches the motif). **B** The relative frequency distribution of orbit 44 for disease proteins (green) and non-disease proteins (red).

**Motif signatures of disease proteins.** The analysis of higher-order PPI network structure can be formalized by counting network motifs, which are subgraphs that recur within a larger network. We here focus on graphlets,[19,20,22,33] connected non-isomorphic induced subgraphs (examples shown in Figure 6). There are 30 possible graphlets of size 2 to 5 nodes. The simplest graphlet is just two nodes connected by an edge, and the most complex graphlet is a clique of size 5. By taking into account the symmetries between nodes in a graphlet, there are 73 different positions or orbits for 2–5-node graphlets, numerated from 0 to 72. For each node in the PPI network we count the number of orbits that the node touches. Motif signature of a protein is thus a set of 73 numbers, $h_i$ ($i = 0, 1, \ldots, 72$) representing the number of induced subgraphs the corresponding node is in, in which the node took the $i$-th orbital position. We use this signature to represent protein's higher-order connectivity in the PPI network.

We conduct permutation tests, comparing the median of the orbit distribution values for proteins associated with a given disease to the medians for 5,000 random samples of sets of proteins of the same size as the disease pathway. These values are used to calculate p-values for each disease at each orbit position. An orbit position for a disease is considered significant if there is an over-representation of counts in the disease proteins compared to the 99% of random samples (*i.e.*, $\alpha = 0.01$).

**Characterization of motifs around disease proteins.** We find that there is a characteristic higher-order PPI network structure around disease proteins (Figure 6), indicating that disease proteins display significance in terms of the orbit positions they tend to inhabit, which could point towards the underlying mechanisms they participate in. We see that 60% (310 of 519) of diseases do show orbit signatures that differ from background proteins and are significantly greater than what one would expect at random. Therefore, even though proteins associated with these diseases may not be adjacent in the PPI network, the diseases do show overall over-representation for certain orbit positions indicating that proteins in disease pathway may take on similar structural roles, albeit in non-adjacent regions in the PPI network.

We can note that orbit position 0 corresponds to subgraphs of two nodes (only an edge), 1 through

Table 1. **Examples of disease-associated motifs.** Shown are 6 orbits (orbit position, *i.e.*, the location where the node touches the motif, is shown in red) whose over-representation is found in most diseases.

| Significant orbit | # of diseases | Examples of diseases with significant orbits |
|---|---|---|
| 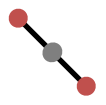 | 78 | Neoplastic cell transformation, Celiac disease, Non-small cell lung carcinoma, Squamous cell carcinoma, Prostatic neoplasms |
|  | 62 | Neoplastic cell transformation, Stomach neoplasms, Restless legs syndrome, Celiac disease, Prostatic neoplasms |
| 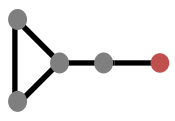 | 62 | Neoplastic cell transformation, IGA glomerulonephritis, Precancerous conditions, Prostatic neoplasms, Liver neoplasms |
| 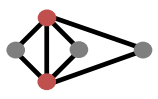 | 88 | Peroxisome biogenesis disorders, Crohn disease, Mitochondrial encephalomyopathies, Venous thromboembolism, Myopia |
| 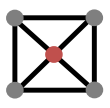 | 229 | Amphetamine-related disorders, Mitochondrial myopathies, Cocaine-related disorders, Nuclear cataract, Polycystic ovary syndrome |
| 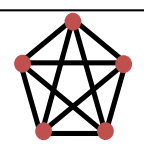 | 228 | Amphetamine-related disorders, Leber congenital amaurosis, Craniofacial abnormalities, Hyperalgesia, Respiratory hypersensitivity |

3 correspond to subgraphs of three nodes, 4 through 14 correspond to subgraphs of four nodes, and the rest correspond to induced subgraphs of five nodes. Therefore, although the distribution of smaller subgraphs such as nodes and triangles are not significant in many diseases, almost 50% of diseases have disease pathways that contain proteins that are over-represented for the most complex orbit positions. For example, orbit position 14 is statistically significantly over-represented in approximately 50 diseases, and position 72 is in over 200 of the diseases (Figure 6, examples in Table 1). We note that we also statistically test for under-representation of the orbit counts, no statistically significant results are observed.

**Characterization of motifs for disease categories.** We also want to investigate whether the orbit signatures are characteristic of diseases in general, or whether there are also differences that could be attributed back to the category the disease belongs to. In order to test this, for each of the 73 orbit positions and for each disease category, we find the statistical significance of the difference in distributions using a two-sample Kolmogorov-Smirnov test. The first sample consists of all the orbit counts for the proteins that are found in at least one disease in a given category, and the other sample consists of all the orbit counts for the proteins that are not associated with any disease in the category. After applying the Bonferroni correction, we then consider a p-value of $\alpha = 0.01$ to be significant.

We find that the most significant differences tend to occur in the more complex motifs. Table 2 shows orbit positions that are considered most characteristic for each disease category, and the graphlets they appear in, which could indicate inherent differences in the manifestations of different classes of diseases.

## 6. Prediction of disease proteins using higher-order PPI network structure

Higher-order PPI network structure is generally not taken into account in current disease protein discovery, although, as showed in Section 5, this structure does encode distinct information about disease proteins. Earlier work showed that motif signatures provide useful signal for biological function prediction,[22,33] but here we want to examine whether they provide additional signal past what edge connectivity in the PPI network already contributes, and if they specifically work for disease protein discovery task.

**Setup and results.** We conduct a logistic regression experiment in which we augment the Neural embeddings (Section 4.2) with information about motif signatures (Section 5). In particular, for each protein we concatenate its neural embedding with its motif signature. Instead of using the motif signature directly, we concatenate the embedding with $h'$, where $h'_i = \max(0, \log(h_i))$, for $i = 0, 1, \ldots, 72$.

Table 2. **Characteristic motifs for disease categories.** Shown are 5 orbits whose over-representation is found in most diseases belonging to a disease category. The orbit position of a node is marked in red.

| Disease category | Significant orbits | Orbit positions |
|---|---|---|
| **Urinary system diseases** *e.g.*, Hyperhomocysteinemia, Nephrosis | 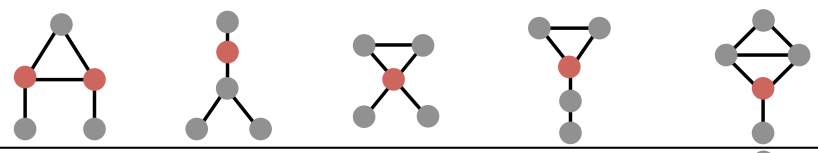 | 26, 20, 33, 30, 47 |
| **Acquired metabolic diseases** *e.g.*, Methylmalonic acidemia, Hyperinsulinism | 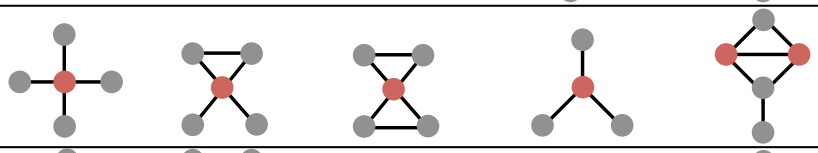 | 23, 33, 44, 7, 48 |
| **Monogenic diseases** *e.g.*, Marfan syndrome, Bardet-Biedl syndrome | 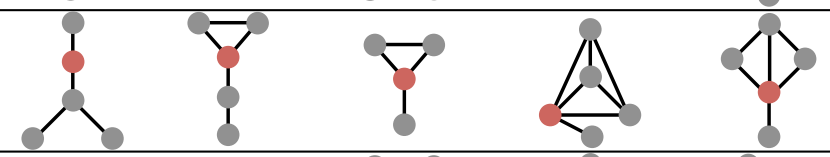 | 20, 30, 11, 42, 58 |
| **Cancer** *e.g.*, Tumor of salivary gland, Papillary thyroid carcinoma | 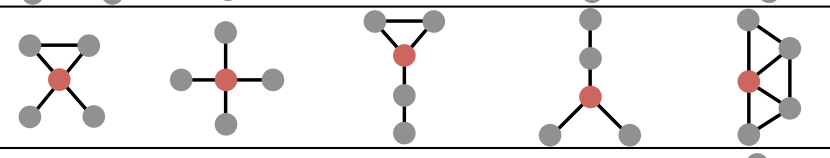 | 33, 23, 30, 21, 61 |
| **Gastrointestinal system diseases** *e.g.*, Eosinophilic esophagitis, Oral fibrosis | 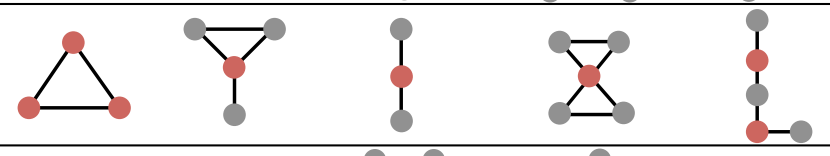 | 3, 11, 2, 44, 16 |
| **Inherited metabolic disorders** *e.g.*, Leigh disease, Mitochondrial complex deficiency | 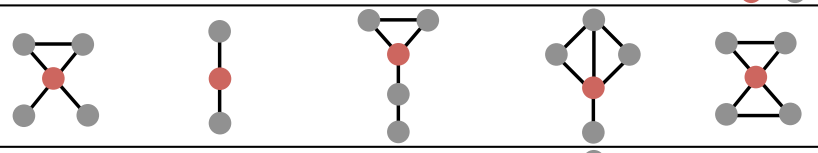 | 33, 2, 30, 42, 44 |
| **Immune system diseases** *e.g.*, Deficiency syndromes, Hypersensitivity | 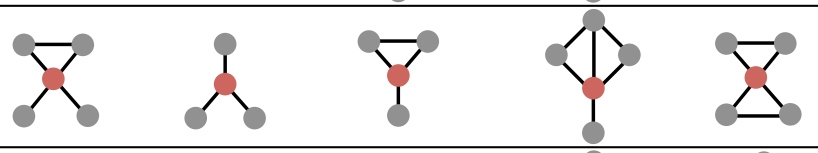 | 33, 7, 11, 42, 44 |
| **Musculoskeletal system diseases** *e.g.*, Muscular atrophy, Muscular dystrophy | 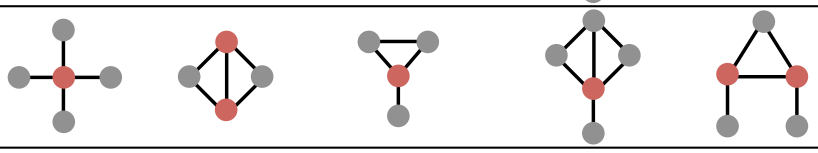 | 23, 13, 11, 42, 26 |
| **Nervous system diseases** *e.g.*, Peripheral neuropathy, Nerve degeneration | 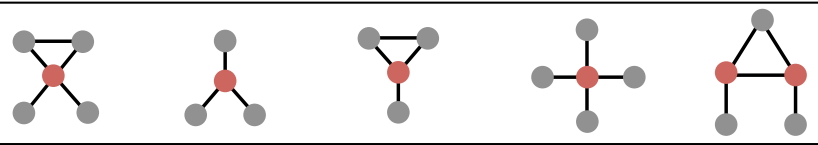 | 33, 7, 11, 23, 26 |
| **Cardiovascular system diseases** *e.g.*, Dilated cardiomyopathy, Tachycardia | 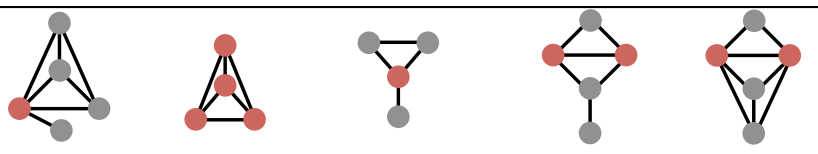 | 58, 14, 11, 48, 67 |

We find that neural embeddings augmented by motif signatures performed on average 11% better than neural embeddings alone (Recall-at-100 = 0.332, compared to Neural embeddings' 0.300). For example, in the case of Hearing loss, the disease that has the greatest increase in performance after the inclusion of higher-order structure, we observe that the recall-at-100 jumps from 0.03 to 0.77 (and the recall-at-100 is at most 0.10 for the four other prediction methods in Section 4.2). If we examine the signature of Hearing loss, as calculated in Section 5, we see that the Hearing loss pathway is significant across all 73 orbit positions, meaning it has a particularly unique signature compared to the background distribution. Though such improvement in performance is not typical across all diseases, this analysis identifies the opportunity to systemically identify diseases which are likely to benefit the most from the inclusion of higher-order PPI network information.

## 7. Conclusion

The overall goal of network biology is to develop approaches that use genomic and other network information to better understand human disease. Given the complexity of this goal, we focused on studying the PPI network structure of disease pathways, defined through sets of proteins associated with diseases. We found that disease pathways are fragmented and sparsely embedded in the PPI network, and that spatial clustering of disease pathways within the PPI network is statistically insignificant. To better understand broad caveats of current methodology for disease protein discovery we evaluated the performance of leading methods and found that their assumptions do not fully capture PPI network structure. We showed, however, that there is detectable higher-order PPI network structure around disease proteins that can be leveraged to boost algorithm performance. These findings provide new insights into the disease pathway PPI network structure and can guide methodological advances in disease protein discovery.

## Acknowledgments

This research has been supported in part by NSF IIS-1149837, NIH BD2K, DARPA SIMPLEX, Stanford Data Science Initiative, and Chan Zuckerberg Biohub.

## References

1. M. D. Ritchie, E. R. Holzinger, R. Li, S. A. Pendergrass and D. Kim, *Nature Reviews Genetics* **16**, 85 (2015).
2. J. Piñero *et al.*, *Database* **2015** (2015).
3. M. Gustafsson *et al.*, *Genome Medicine* **6**, p. 82 (2014).
4. P. Creixell *et al.*, *Nature Methods* **12**, p. 615 (2015).
5. S. Navlakha and C. Kingsford, *Bioinformatics* **26**, 1057 (2010).
6. N. Natarajan and I. S. Dhillon, *Bioinformatics* **30**, i60 (2014).
7. S. D. Ghiassian, J. Menche and A.-L. Barabási, *PLoS Computational Biology* **11**, p. e1004120 (2015).
8. H. Zhou and J. Skolnick, *Bioinformatics* **32**, 2831 (2016).
9. Y. Silberberg, M. Kupiec and R. Sharan, *Genome Medicine* **9**, p. 48 (2017).
10. S. van Dam *et al.*, *Briefings in Bioinformatics* **bbw139**, 1 (2017).
11. A. Sharma *et al.*, *Human Molecular Genetics* (2015).
12. A. Krishnan, J. N. Taroni and C. S. Greene, *Current Genetic Medicine Reports* **4**, 155 (2016).
13. J. Loscalzo and A.-L. Barabási, *Wiley Interdisciplinary Reviews: Systems Biology and Medicine* **3**, 619 (2011).
14. A.-L. Barabási, N. Gulbahce and J. Loscalzo, *Nature Reviews Genetics* **12**, 56 (2011).
15. L. I. Furlong, *Trends in Genetics* **29**, 150 (2013).
16. M. Zitnik and B. Zupan, *Bioinformatics* **32**, 90 (2016).
17. M. Zitnik and B. Zupan, *Pacific Symposium on Biocomputing* **21**, p. 81 (2016).
18. J. Menche *et al.*, *Science* **347**, p. 1257601 (2015).
19. N. Pržulj, D. G. Corneil and I. Jurisica, *Bioinformatics* **22**, 974 (2006).
20. N. Pržulj, *Bioinformatics* **23**, 177 (2007).
21. K. Sun, J. P. Gonçalves, C. Larminie and N. Pržulj, *BMC Bioinformatics* **15**, p. 304 (2014).
22. T. Milenković *et al.*, *Journal of the Royal Society Interface* **7**, 423 (2010).
23. M. D. Leiserson *et al.*, *Nature Genetics* **47**, 106 (2015).
24. A. Grover and J. Leskovec, *ACM SIGKDD* **22**, 855 (2016).
25. A. Chatr-Aryamontri *et al.*, *Nucleic Acids Research* **43**, D470 (2015).
26. D. Szklarczyk *et al.*, *Nucleic Acids Research* **43**, 447 (2015).
27. W. A. Kibbe *et al.*, *Nucleic Acids Research* **43**, D1071 (2014).
28. S. E. Schaeffer, *Computer Science Review* **1**, 27 (2007).
29. A. J. Cornish and F. Markowetz, *PLoS Computational Biology* **10**, p. e1003808 (2014).
30. A. Baryshnikova, *Cell Systems* **2**, 412 (2016).
31. M. E. Newman, *Proceedings of the National Academy of Sciences* **103**, 8577 (2006).
32. J. Loscalzo, I. Kohane and A.-L. Barabási, *Molecular Systems Biology* **3**, p. 124 (2007).
33. W. Hayes, K. Sun and N. Pržulj, *Bioinformatics* **29**, 483 (2013).